\documentclass{article}
\usepackage{dcase2019,cite,url}
\usepackage{graphicx,amsmath,amssymb,multirow,dblfloatfix,siunitx, bbold,url}

\title{A dataset of reverberant spatial sound scenes with moving sources for sound event localization and detection}

\name{Archontis Politis, Sharath Adavanne, and Tuomas Virtanen\thanks{This work has received funding from the European Research Council under the ERC Grant Agreement 637422 EVERYSOUND.}}
\address{Audio and Speech Processing Research Group, Tampere University, Finland}

\begin{document}
\ninept
\maketitle

\setlength{\abovedisplayskip}{5pt}%
\setlength{\belowdisplayskip}{5pt}%
\setlength{\abovedisplayshortskip}{5pt}%
\setlength{\belowdisplayshortskip}{5pt}%

\begin{abstract}
This report presents the dataset and the evaluation setup of the Sound Event Localization \& Detection (SELD) task for the DCASE 2020 Challenge. The SELD task refers to the problem of trying to simultaneously classify a known set of sound event classes, detect their temporal activations, and estimate their spatial directions or locations while they are active. To train and test SELD systems, datasets of diverse sound events occurring under realistic acoustic conditions are needed. Compared to the previous challenge, a significantly more complex dataset was created for DCASE 2020. The two key differences are a more diverse range of acoustical conditions, and dynamic conditions, i.e. moving sources. The spatial sound scenes are created using real room impulse responses captured in a continuous manner with a slowly moving excitation source. Both static and moving sound events are synthesized from them. Ambient noise recorded on location is added to complete the generation of scene recordings. A baseline SELD method accompanies the dataset, based on a convolutional recurrent neural network, to provide benchmark scores for the task. The baseline is an updated version of the one used in the previous challenge, with input features and training modifications to improve its performance.
\end{abstract}


\section{Introduction}

Sound event localization and detection (SELD) takes the currently active research topic of temporal sound event detection (SED) \cite{mesaros2019sound} and connects it with the spatial dimension of event location or acoustic direction-of-arrival (DoA). Hence SELD aims to a more complete spatiotemporal characterization of the acoustic sound scene, with predictions on the type of sounds of interest in the scene, their temporal activations, and their spatial trajectories when they are active. This spatiotemporal scene description has a wide range of applications in machine listening, ranging from acoustic monitoring and robot navigation to intelligent human-machine interaction and deployment of immersive services.

Until the DCASE2019 Challenge\footnote{\url{http://dcase.community/challenge2019/task-sound-event-localization-and-detection}}, only a handful of approaches in literature were aiming some form of SELD~\cite{valenzise2007scream, butko2011two, chakraborty2014sound, Hirvonen2015, Lopatka2016, grobler2017sound, Adavanne_JSTSP2018}. Apart from \cite{Hirvonen2015, Adavanne_JSTSP2018} which are fully deep-neural network (DNN) based approaches, these earlier works employed more traditional source localization methods such as time-difference-of-arrival (TDoA) \cite{valenzise2007scream, grobler2017sound}, steered-response power \cite{butko2011two}, or acoustic intensity vector analysis \cite{Lopatka2016}, and Gaussian mixture models \cite{valenzise2007scream}, hidden Markov models \cite{butko2011two}, support vector machines \cite{Lopatka2016}, or a simple artificial neural network \cite{grobler2017sound} for classification. Additionally, most of them treated detection and localization independently, with only \cite{grobler2017sound, chakraborty2014sound} joining beamforming outputs after localization with the event classifiers.

Recently, DNNs have dominated SED approaches \cite{mesaros2019sound}, and they have been applied successfully to pure source localization \cite{Adavanne2018_EUSIPCO, perotin2019crnn, chakrabarty2019multi}, showing potential for joint modeling of the SELD task. The first works we are aware of this approach are~\cite{Hirvonen2015, Adavanne_JSTSP2018}. Hirvonen in~\cite{Hirvonen2015} used a convolutional neural network (CNN) with localization targets at discrete directions-of-arrival (DoAs), setting the SELD task as a multilabel-multiclass classification problem. In \cite{Adavanne_JSTSP2018} we proposed the SELDnet, a convolutional recurrent neural network (CRNN) with two output branches, one for SED and the other for localization. Contrary to \cite{Hirvonen2015}, localization here was based on a regression approach with one DoA predicted per sound class. Both proposals were using simple generic features, such as multichannel power \cite{Hirvonen2015}, or phase and magnitude \cite{Adavanne_JSTSP2018}, spectrograms.

Due to its relevance in all the aforementioned applications, SELD was introduced as a new task in DCASE 2019 Challenge, and as such, it required a new dataset for training and evaluation of the submitted methods. This dataset, the \textbf{TAU Spatial Sound Events 2019}\footnote{\url{https://zenodo.org/record/2580091}}, comprised scenes with events from 11 classes, spatialized through captured room impulse responses (RIRs) as static sources at 504 possible locations for each of 5 different spaces \cite{Adavanne2019a}. Along with the dataset, a SELDnet implementation was provided by the authors as a baseline for the challenge participants\footnote{\url{https://github.com/sharathadavanne/seld-dcase2019}}. The challenge attracted more than 20 original methods, with most methods surpassing significantly the baseline\footnote{\url{http://dcase.community/challenge2019/task-sound-event-localization-and-detection-results}}. Many innovative solutions were presented for the task, such as more refined SED and localization features \cite{Cao2019, Grondin2019, Cordourier2019}, a multi-stage modeling and training approach \cite{Cao2019}, data augmentation \cite{Mazzon2019, Pratik2019}, exploitation of domain-specific knowledge \cite{Kapka2019, Grondin2019}, state-of-the-art network architectures \cite{Ranjan2019, Park2019a}, ensembles \cite{Chytas2019}, or combinations of model-based localization and DNN-based event detection \cite{Perez-Lopez2019}.

In this work we present the new dataset \textbf{TAU-NIGENS Spatial Sound Events 2020}\footnote{\url{https://zenodo.org/record/3740236}} aimed for the next iteration of the SELD task in DCASE 2020 challenge\footnote{\url{http://dcase.community/challenge2020/task-sound-event-localization-and-detection}}. The dataset preserves all the realistic properties of the previous one: realistic reverberation and ambient noise based on real measured spaces, variable acoustic conditions from a variety of rooms, large range of source positions with respect to the microphone, and two different recording array formats for the participants to exploit. However, the dataset overcomes the major limitations of the past one: more sound examples per class, a greater number of rooms, much more diverse acoustic conditions, and non-quantized source positions in a predefined grid of directions. Apart from improvements, the dataset introduces moving sources for about half of the active events, which makes it significantly more challenging, demanding closer to a real-life performance from the submissions.

Along with the dataset, we introduce improvements on two additional fronts. Firstly, the baseline implemented and published along with the dataset remains the SELDnet architecture of \cite{Adavanne_JSTSP2018}, due to its conceptual simplicity and efficient architecture. However, several small changes are introduced that reflect the most common improvements used by DCASE2019 participants, to make it more effective with the new more challenging dataset. Secondly, instead of measuring performance independently for SED and localization, as in DCASE2019, we introduce the recently proposed metrics that consider joint SELD performance \cite{mesaros2019joint}, reflecting better the expected performance differences between systems that localize the correct events at their correct position, and systems that detect and/or localize well independently.

\section{Reverberant dynamic dataset}

\subsection{Sound events}

Sound event samples were sourced from the recently published NIGENS General Sound Events Database\footnote{\url{https://zenodo.org/record/2535878}}. This database provides a higher number of samples and classes than the one used in the previous challenge. 714 sound examples are distributed between 14 classes of \emph{alarm, crying baby, crash, barking dog, running engine, burning fire, footsteps, knocking on door, female \& male speech, female \& male scream, ringing phone, piano}. For more details on the recordings and the database in general, the reader is referred to \cite{trowitzsch2019nigens}.

\subsection{Recording of multichannel RIRs}

The overall recording procedure was kept similar to the one employed in the previous dataset \cite{Adavanne2019a}, with differences highlighted below. For DCASE2019, real recorded RIRs were captured from 5 rooms. However, all of those rooms were large publicly accessible open spaces in university buildings. Furthermore, the grid of source positions around the microphone was kept constant between rooms, with two possible source distances at 1m or 2m. Both of these conditions resulted in the direct sound and floor reflections being dominant and, in general, in high direct-to-reverberant ratios (DRR). For DCASE2020, to add more variability in acoustical conditions and more challenging reverberation, we recorded 10 more rooms of diverse shapes and types, such as lecture halls, large classrooms, small classrooms and meeting rooms, a modern sports hall, and a sports hall in an underground nuclear shelter with rock walls. 

Similarly as in the DCASE2019 dataset, instead of RIR measurements at discrete source-receiver points, a very large range of source positions is covered by recording pseudo-random noise (MLS) emitted by a slowly moving source along predefined tracks \cite{hahn2016comparison}. The source is a Genelec G Three\footnote{\url{https://www.genelec.com/g-three}} loudspeaker mounted on a wheeled platform. The platform is moved manually during the duration of the recording, while the microphone array is immobile. The recording is done with a 32-channel compact spherical microphone array (SMA), the em32 Eigenmike\footnote{\url{https://mhacoustics.com/products#eigenmike1}}. An SMA with high channel count is chosen due to its uniform spatial resolution up to high frequencies, and to its flexibility in allowing us to extract a variety of smaller spatial formats from the same recording.

Contrary to the DCASE2019 dataset, the recording trajectories in the new rooms are different for each one of them. In some rooms, the recordings were done in circular trajectories, but at differing distances and elevations, while in other rooms linear trajectories at various heights were used. The RIRs were extracted from the moving source recordings through a simple linear regression on the filter coefficients between the clean MLS sequence and the recorded output, similar to \cite{avargel2007multiplicative}. RIRs extracted along circular trajectories have a more or less constant elevation, distance, and DRR, while ones extracted along linear trajectories have varying elevation, distance, and DRR, with respect to the recording position.

Similarly to DCASE2019, apart from the MLS noise sequences, 30 mins of spatial ambient noise were additionally captured in each room with the recording setup unchanged. Contrary to the 5 earlier rooms which were accessible by passing crowds at any time, the new room recordings contained mostly ventilation noise.

\subsection{Reference RIRs and positional labels}

During the synthesis of the spatial mixtures, sound events are intended to be spatialized at consistent DoAs across different environments, meaning that the direct path for the same DoA, as encoded in the array channels, should be similar between rooms so that the methods can rely on it for localization while being robust to the dissimilar reverberation patterns that follow. In the DCASE2019 dataset, the recorded trajectories were assumed to have the exact same geometry with respect to the microphones, across rooms. Static RIRs were extracted along circular trajectories with an angular separation of 10 degrees, and the final grid of reference positions was intended to be the same for all rooms. In this case, we found good alignment between the intended geometric positions and the actual acoustic DoAs seen by the microphone array, and the reference DoA-RIR pairs were assumed to be on a spherical grid of fixed azimuths and elevations.

Assuming a constant speed of motion, the same process could have been applied to the new more spatially complex measurements, since the geometry of each trajectory was planned beforehand. However, this assumption proved unrealistic, due to varying speeds and geometric misalignment between the intended and the real recording geometry. To address this issue, an additional 360 video track was recorded along with the audio recording, with the camera 10 cm above the microphone array, and a simple video object tracking algorithm, bounded on the loudspeaker, was used for estimating a reference DoA at all times of the recording. However, even though the video tracking was stable, it was found that small rotational misalignment between the camera and the array frame of reference could reflect large DoA differences.

According to the above, we finally decided to estimate the reference DoAs acoustically, directly from the extracted RIRs, as these would reflect consistently the ones encoded into the multichannel mixture during the synthesis stage using the same RIRs. To that purpose, for each source trajectory we: a) extracted the multichannel RIRs at 200-millisecond intervals, b) estimated direct path delays from geometry and measurements, c) windowed the RIRs around their direct path, and d) applied a broadband version of the subspace MUSIC algorithm for estimation of the DoA corresponding to that early part of the RIR. From that list of RIR-DoA pairs, the final reference ones were determined by selecting the ones closest to the geometric reference trajectory, at approximately 1-degree intervals. Note that the same process was applied also to the 5 earlier rooms recorded for the DCASE2019 dataset.

\subsection{Dataset Synthesis}
All extracted multichannel RIRs and sound event samples were resampled to 24 kHz. From the 8 provided splits of the NIGENS dataset, 6 were used for the creation of the development, and the remaining 2 for the evaluation datasets. One or two rooms were assigned to each split, and 100 mixtures of spatialized sound events were generated for each such combination of event samples and rooms. Each generated mixture was 1 minute long. The onsets of sound events in each recording were randomly distributed but constrained by the allowed level of polyphony (number of simultaneous events), which could be either one or two.

An event was randomly chosen to be either static or moving. Static events were assigned randomly a DoA from the list of reference ones available for the specific room. Moving sound events were assigned randomly one of the RIR recording trajectories for the specific room, hence limiting their motion along that path. However, the movement direction and the rate of motion could be different for each event. The direction of movement was randomized, while the speed of motion was randomly chosen from three levels of \emph{slow} ($\sim$10 deg/sec), \emph{medium} ($\sim$20 deg/sec), and \emph{fast} ($\sim$40 deg/sec). Additionally, since each trajectory was recorded at different heights, moving sound events reaching the end of a path had the possibility to jump to a higher or lower elevation and continue their motion on the respective path of that height.

Static events were spatialized by convolution with the respective RIRs for their intended DoA, and added to the mixture. Moving sound events were spatialized by a time-variant convolution scheme, performed between the STFT of the event sample and the STFTs of all the RIRs encountered along the path of motion. The operation resembled a partitioned convolution scheme, with RIRs being combined with a cross-fading scheme giving more weight to frames of past RIRs for the reverberation tail, and more weight to frames of the recently encountered RIRs for the direct path and early reflections. Since the reference DoAs were extracted at about 1$^{\circ}$ intervals along a trajectory, the speed of motion was controlled by using 10 (slow), 20 (medium), or 40 (fast) consecutive RIRs per 1 second of output. Very short events of up to 2 seconds were excluded from being dynamic, and were assigned static DoAs instead.

After the convolved spatialized sound events were added to each multichannel mixture with the intended polyphony, ambient noise from the same room was additionally mixed. The original ambient noise recordings were split into 1-minute segments and added to the mixtures at varying signal-to-noise (SNR) levels between from 30 dB to 6 dB. An omnidirectional component was extracted through a linear combination of the channels of the noiseless mixture and the ambient noise recording, and the power ratio between the two signals was tuned to match the intended SNR. The respective gain factor was then applied to the ambient noise segment before adding to the mixture. Since the duration of the recorded ambient noise at each room was less than the total duration of the mixtures generated for that room, additional 1-minute noise segments were artificially generated by simply mixing two randomly chosen segments of the recording.

\subsection{Dataset Formats}

As in the previous dataset for DCASE2019, we opted for delivering the synthesized sound recordings in two different 4-channel spatial sound formats, extracted from the 32-channel Eigenmike format. The first format is a 4-channel microphone array one, extracted directly by selecting a subset of the Eigenmike channels, corresponding to a tetrahedral capsule arrangement (MIC). The second format is the widespread first-order Ambisonics (FOA), extracted through a matrix of $4\times 32$ conversion filters, as detailed in \cite{politis2017comparing}. The rationale behind offering the dataset in both the MIC and FOA formats is that each one encodes spatial information differently. The MIC array format has microphones arranged in spherical coordinates of ($\ang{45}$, $\ang{35}$, 4.2 cm), ($\ang{-45}$, $\ang{-35}$, 4.2 cm), ($\ang{135}$, $\ang{-35}$, 4.2 cm) and ($\ang{-135}$, $\ang{35}$, 4.2 cm), taken from channels 6, 10, 26, and 22 of the Eigenmike, encoding a DoA with both time-differences, due to the spacing, and level differences, due to the acoustical shadowing of the hard spherical baffle in between. On the other hand, the FOA format is space-coincident, offering only level differences and no time-differences for a single DoA. Hence different features spatial features are better suited to each format, and participants could exploit one of the two or both.

For model-based and parametric localization approaches, the multichannel response with respect to a given source DoA, should be known. The spatial responses of the MIC and FOA formats were described in \cite{Adavanne2019a} and are repeated here for the sake of completeness. The directional responses of the $m$th channel $H_m(\phi,\theta,f)$ to a source incident from DOA given by azimuth angle $\phi$ and elevation angle $\theta$, at frequency $f$, is for the FOA format:
\begin{eqnarray}
H_1(\phi, \theta, f) &=& 1 \\
H_2(\phi, \theta, f) &=& \sin(\phi) * \cos(\theta) \\
H_3(\phi, \theta, f) &=& \sin(\theta) \\
H_4(\phi, \theta, f) &=& \cos(\phi) * \cos(\theta),
\end{eqnarray}
corresponding to the \emph{SN3D} normalization scheme of Ambisonics. The format is assumed frequency-independent, which holds true up to about 9 kHz for FOA encoded from the Eigenmike, and deviates gradually from the ideal response for higher frequencies.

For the tetrahedral array of microphones mounted on a spherical baffle, an analytical expression for the directional array response is given by the expansion:
\begin{align}
& H_m(\phi_m, \theta_m, \phi, \theta, \omega) = \nonumber\\ 
& \frac{1}{(\omega R/c)^2}\sum_{n=0}^{30} \frac{\mathrm{i}^{n-1}}{h_n'^{(2)}(\omega R/c)}(2n+1)P_n(\cos\gamma_m),
\end{align}
where $m$ is the channel number, $(\phi_m, \theta_m)$ are the specific microphone's azimuth and elevation position, $\omega = 2\pi f$ is the angular frequency, $R = 0.042$ m is the array radius, $c = 343$ m/s is the speed of sound, $\cos(\gamma_m)$ is the cosine angle between the microphone position and the DOA, $P_n$ is the unnormalized Legendre polynomial of degree $n$, and $h_n'^{(2)}$ is the derivative with respect to the argument of a spherical Hankel function of the second kind. The expansion is limited to 30 terms which provide negligible modeling error up to 20 kHz.







\begin{figure}[t]
  \centering
  \centerline{\includegraphics[width=0.49\textwidth ,keepaspectratio]{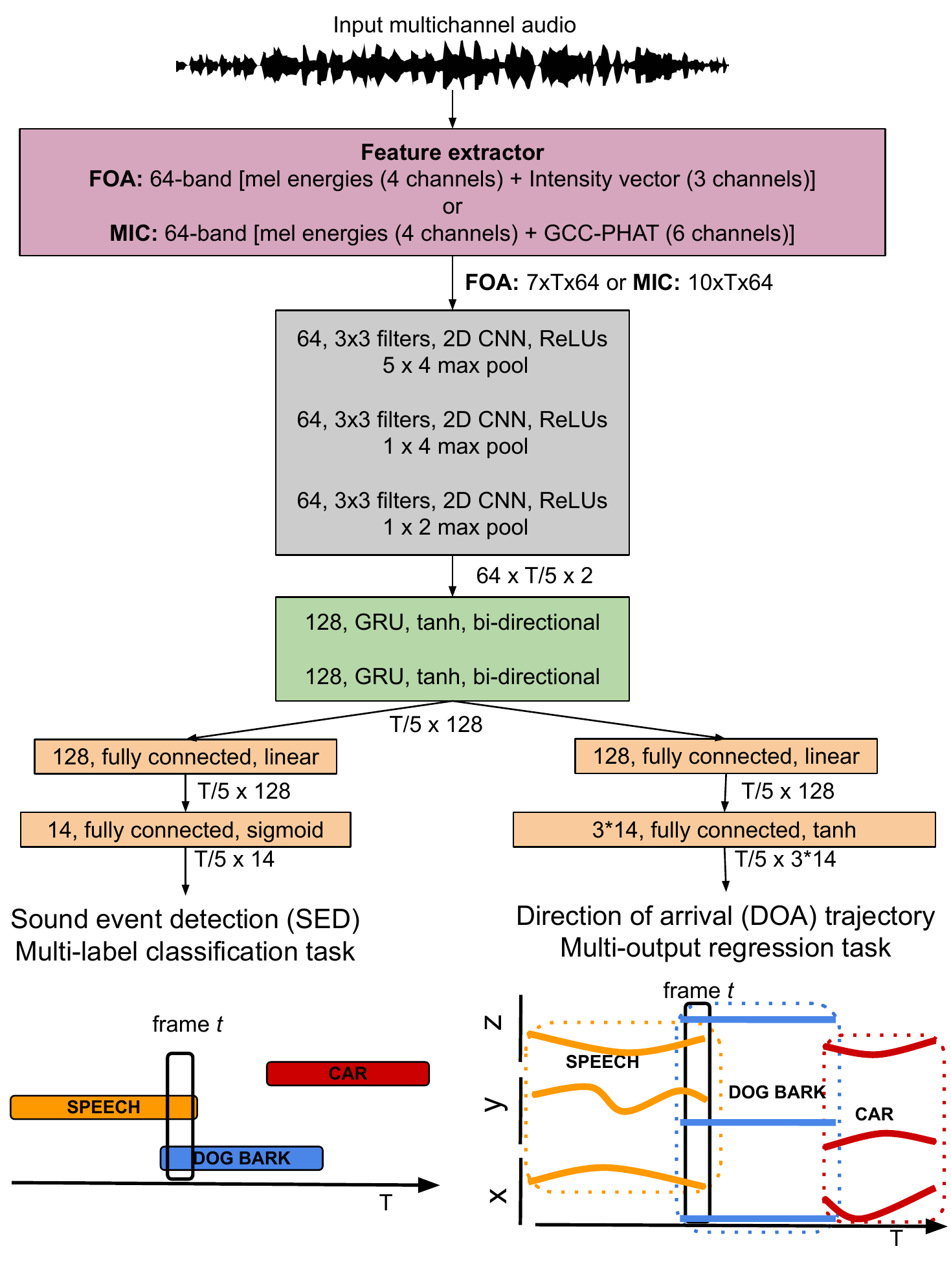}} 
  \caption{Convolutional recurrent neural network for SELD.}
  \label{fig:crnn}
\end{figure}

\section{Baseline Method}
As the benchmark method, we employ an updated version of  SELDnet~\cite{Adavanne_JSTSP2018}\footnote{\url{https://github.com/sharathadavanne/seld-dcase2020}}. Specifically, we adopt changes to SELDnet that helped to improve its performance consistently across different submissions of the DCASE 2019 SELD task. The general improvements proposed in the DCASE 2019 task submissions broadly fall into four categories - a) array-dependent acoustic feature extraction to enhance learning \cite{Cao2019, Grondin2019}, b) different deep-learning architectures, including separate models for SED and DOA estimation for robust learning \cite{Cao2019, Ranjan2019, Park2019a}, c) improved training objective for DOA estimation, by employing SED reference as a mask \cite{Cao2019}, and d) post-processing of the SELD output based on the dataset characteristics. Among these improvements, for the current baseline method we include array-dependent acoustic feature extraction, and train a single model to jointly estimate SED and DOA as shown in Figure~\ref{fig:crnn}. Additionally, during the training, the DOA estimation branch uses the SED output as the mask, and the mean squared error loss is only computed for the sound events that are active. This strategy was first published by \cite{Cao2019}, with significant improvements on the results, and adopted by other participants in the challenge. Similar to the original SELDnet, we do not perform any post-processing on its output.

The updated SELDnet takes as input multichannel audio at 24 kHz sampling rate. For each of the two datasets, MIC and FOA, two features are extracted. The first feature, the multichannel mel-band power spectrogram, is common to both datasets, and, apart from being a popular SED feature, it additionally captures inter-channel level differences (ILDs). It is computed for each channel as 64 log mel-band energies with a 40 ms window, and 20 ms hop length using a 1024-point FFT. The second, format-specific, spatial feature for the FOA dataset is the acoustic intensity vector, which expresses net acoustic energy flux, and is computed at each of the 64 mel-bands similar to~\cite{Cao2019, Park2019a}. For the MIC dataset, we employ the generalized-cross-correlation with phase-transform (GCC-PHAT) feature computed in each of the 64 mel-bands similar to~\cite{Cao2019, Pratik2019, Mazzon2019}.

Based on the chosen dataset, the SELDnet is trained using the corresponding features. For the FOA dataset, the input is of $7\times T\times64$ dimension, where $T$ is the number of time frames in the input sequence, and the number $7$ arises from $4$ channels of $64$ dimension log mel-band energies computed for each of the $4$ audio-channels, and $3$ channels of FOA intensity vectors. Similarly for the MIC dataset, the input is of $10\times T\times64$, where $10$ arises from $6$ channels of GCC-PHAT computed between all pairs of audio-channels of the MIC dataset and $4$ channels of log mel-band energies.

Irrespective of the dataset, we employ three convolutional layers to learn shift-invariant features from the input acoustic feature. Both the temporal and frequency resolution of the input is reduced using a max-pooling operation after every convolutional layer. The final temporal resolution is equal to 100 ms, which is the one specified by the challenge submission format. Two layers of gated recurrent units are employed to learn the temporal structure from the convolutional features. Thereafter separate branches of fully-connected layers are employed to learn SED and DOA estimation. The SED output layer has \textit{sigmoid} activations and generates an output of the dimension $T/5 \times C$, which corresponds to the temporal activity of the $C$ classes ($C$ = 14) at 100 ms resolution. Similarly, the DOA output layer has \textit{tanh} activations and generates an output of the dimension $T/5 \times 3C$, which corresponds to the DOA trajectory of the $C$ classes at the same temporal resolution. The value $3C$ is due to the Cartesian representation of the DOA for each of the $C$ classes. During training, the SED branch uses the binary cross-entropy objective, whereas the DOA branch is updated to use the masked MSE loss discussed above. The updated SELDnet is trained using Adam optimizer with a weighted combination of SED and DOA loss, where DOA loss is weighed 1000 times more than SED loss.

\begin{table}[t]
\centering  
\renewcommand\thetable{1}
\caption{Evaluation setup}
\begin{tabular}{l|ccc}
 & \multicolumn{3}{c}{\textbf{Splits}} \\ \cline{2-4}
\textbf{Dataset} & \textbf{Training} & \textbf{Validation} & \textbf{Testing} \\ \hline
\textbf{Development}  & 3, 4, 5, 6 & 2 & 1 \\
\end{tabular}
\label{T:splits}
\end{table}

\section{Evaluation}
\subsection{Evaluation Setup}
The evaluation setup for the development dataset is shown in Table~\ref{T:splits}. Among the six splits in the dataset, the first set is used as the unseen test split, the second set is used as the validation split during training, and the remaining sets are used for training. The correct usage of this evaluation setup is as follows. The best parameters for a SELD method are chosen based on the validation split, without using the testing split. The performance of the best validation model on the unseen testing split is then reported as the development dataset score for the SELD method. The baseline SELDnet method chooses the best validation model based on early stopping on the validation split. Thereafter, the SELD performance on the unseen test data is computed using the best validation model.

In order to have a fair comparison of the SELD performance across different submitted systems, participants are required to employ the proposed evaluation setup and report the performance of their method on the unseen test split. However, for the evaluation dataset, the participants are allowed to decide the training procedure, i.e. the amount of training and validation files in the development dataset and the number of ensemble models.

\begin{figure}[t]
  \centering
  \centerline{\includegraphics[width=0.49\textwidth ,keepaspectratio, trim=0.1cm 0.1cm 0.1cm 0.1cm,clip]{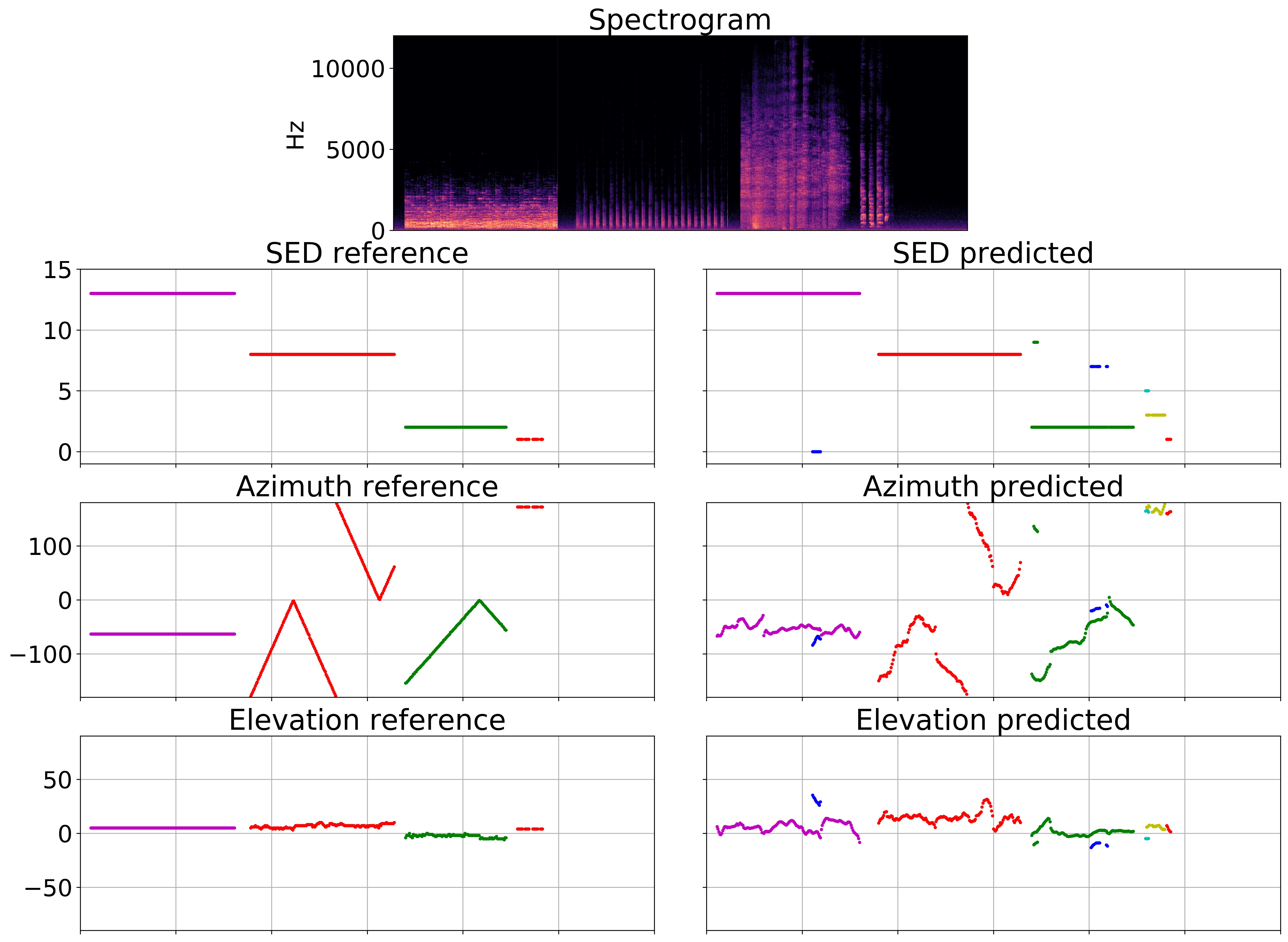}} 
  \caption{Visualization of SELD performance for baseline method.}
  \label{fig:crnn-output}
\end{figure}

\subsection{Metrics}
\label{ssec:metrics}
The 2019 version of the SELD task employed individual metrics for SED and DOA estimation. The SED performance was evaluated using the F-score ($F$) and error rate ($ER$) calculated in non-overlapping one-second segments~\cite{metrics}. The DOA estimation was evaluated using frame-wise metrics~\cite{Adavanne2018_EUSIPCO} of DOA error ($DE$) and frame recall ($FR$). The DOA error represents the average angular error in degrees between the predicted and reference DOAs. The frame recall represents the percentage of frames in which the estimated number of DOAs were identical to the reference.

Recently, in~\cite{mesaros2019joint} we discussed the drawbacks of the above metrics for the SELD task and proposed metrics to evaluate the performance of joint detection and localization. The first two metrics, on location-aware detection, consider a prediction to be correct if the sound class of the prediction and reference are the same, and the distance between them is below a threshold. For the SELD task we propose to use a threshold of $20^\circ{}$, and compute the corresponding metrics - error rate ($ER_{20^\circ{}}$) and F-score ($F_{20^\circ{}}$) in one-second non-overlapping segments. An ideal SELD method will have $ER_{20^\circ{}} = 0$ and $F_{20^\circ{}} = 100\%$.

The next two metrics, on class-aware localization, do not use any distance threshold, like above, but consider the error only between same-class predictions and references. The respective localization error ($LE_{CD}$) and its corresponding localization recall ($LR_{CD}$) are computed in one-second non-overlapping segments, where the subscript refers to classification-dependent. An ideal SELD method will have $LE_{CD}=0^\circ$ and $LR_{CD}$ of 100\%.

Although the information on joint localization/detection performance can be gained by either location-aware detection, or class-aware localization, a more complete picture is given by all four. Hence, we evaluate all the submissions in the DCASE2020 task using all four metrics. The submitted methods will be ranked individually for each one of them, and the final positions will be obtained using the cumulative minimum of the ranks.

\section{Results}

The input log mel-band spectrum, and the corresponding output of the baseline SELDnet method, for a recording from the unseen test split, is shown in Figure~\ref{fig:crnn-output}. Each sound event class is represented by a unique color across the sub-plots. We observe that the baseline SELDnet method performs joint detection, localization, and tracking of dynamic sources successfully in this case.

The performance of the SELDnet method for the proposed evaluation setup of the DCASE 2020 SELD task is tabulated in Table~\ref{tab:overall-results}. The results for both the DCASE2019 metrics and the official DCASE2020 metrics are reported. The 2020 metrics evaluate jointly detection and localization performance and hence provide deeper insights on the SELD performance. For instance, the 2019 detection metrics of $DE$ and $FR$ suggest that the SELDnet estimated the correct number of DOAs in 66.6\% of the frames for the FOA test data with an average DOA error of 20.4$^\circ{}$. But, this localization metric does not use the knowledge of detection and computes DOA error for all the detected sound classes, irrespective of them being correct or wrong. Although we have the corresponding detection scores of $ER$ and $F$ scores, there is no straight-forward approach to assess a joint detection and localization performance. In contrast, the 2020 metrics of class-aware localization ($LE_{CD}$ and $LR_{CD}$) and location-aware detection ($ER_{20^\circ{}}$) and F-score ($F_{20^\circ{}}$) can both independently provide insights on the joint performance. For instance, on the FOA test data, 60.7\% ($LR_{CD}$) of the sound class instances were recalled successfully by the SELDnet with an average location error ($LE_{CD}$) of 22.8$^\circ{}$. Similarly, if we consider that the predicted sound class is correct if it is within a margin of 20$^\circ{}$ from the reference sound class location, then we obtain an F-score ($F_{20^\circ{}}$) of 37.4\% and error rate ($ER_{20^\circ{}}$) of 0.72.

\begin{table}[ht]
\caption{SELD performance of the baseline method evaluated using independent (2019) and joint (2020) localization/detection metrics.}
\label{tab:overall-results}
\resizebox{\columnwidth}{!}{%
\begin{tabular}{l|cccc|cccc}
 & \multicolumn{4}{c|}{\textbf{FOA}} & \multicolumn{4}{c}{\textbf{MIC}} \\\cline{2-9}
\textbf{2019} & \textbf{$DE$} & \textbf{$FR$} & \textbf{$ER$} & \textbf{$F$} & \textbf{$DE$} & \textbf{$FR$} & \textbf{$ER$} & \textbf{$F$} \\\hline
\multicolumn{9}{l}{\textbf{Development results}} \\ \hline
\textbf{Val} & 20.2$^\circ{}$  & 62.9 & 0.54 & 62 & 21.9$^\circ{}$  & 63.8 & 0.53 & 62.8 \\
\textbf{Test} & 20.4$^\circ{}$  & 66.6 & 0.54 & 60.9 & 22.6$^\circ{}$  & 66.8 & 0.56 & 59.2 \\
\multicolumn{9}{l}{} \\
\textbf{2020} & \textbf{$LE_{CD}$} & \textbf{$LR_{CD}$} & \textbf{$ER_{20^\circ{}}$} & \textbf{$F_{20^\circ{}}$} & \textbf{$LE_{CD}$} & \textbf{$LR_{CD}$} & \textbf{$ER_{20^\circ{}}$} & \textbf{$F_{20^\circ{}}$} \\\hline
\multicolumn{9}{l}{\textbf{Development results}} \\\hline
\textbf{Val} & 23.5$^\circ{}$  & 62 & 0.72 & 37.7 & 27$^\circ{}$  & 62.6 & 0.74 & 34.2 \\
\textbf{Test} & 22.8$^\circ{}$  & 60.7 & 0.72 & 37.4 & 27.3$^\circ{}$  & 59 & 0.78 & 31.4 \\
\multicolumn{9}{l}{\textbf{Detailed development dataset test-split results}} \\\hline
\textbf{Overlap 1} & 18.1$^\circ{}$ & 69.7 & 0.63 & 49.2 & 20.8$^\circ{}$ & 66.6 & 0.70 & 40.8 \\
\textbf{Overlap 2} & 26.3$^\circ{}$ & 55.4 & 0.77 & 30.4 & 32.0$^\circ{}$ & 54.6 & 0.82 & 25.8 \\
\multicolumn{9}{l}{} \\\hline
\multicolumn{9}{l}{\textbf{Evaluation results}} \\\hline
\textbf{Val} & 22.8$^\circ{}$  & 60.7 & 0.7 & 39.6 & 24.5$^\circ{}$  & 58.7 & 0.72 & 36,9 \\
\textbf{Test} & 23.2$^\circ{}$  & 62.1 & 0.7 & 39.5 & 23.1$^\circ{}$  & 62.4 & 0.69 & 41.3 \\
\multicolumn{9}{l}{\textbf{Detailed evaluation dataset test-split results}}\\ \hline
\textbf{Overlap 1} & 18.3$^\circ{}$ & 69.9 & 0.58 & 51.3 & 16.0$^\circ{}$ & 69.4 & 0.75 & 33.7 \\
\textbf{Overlap 2} & 26.7$^\circ{}$ & 57.4 & 0.75 & 32.5 & 28.1$^\circ{}$ & 58.3 & 0.75 & 33.7 \\
\textbf{Split 7} & 20.5$^\circ{}$ & 65.0 & 0.66 & 43.3 & 21.8$^\circ{}$ & 65.9 & 0.66 & 44.0 \\
\textbf{Split 8} & 26.2$^\circ{}$ & 59.1 & 0.74 & 35.5 & 24.7$^\circ{}$ & 58.9 & 0.72 & 38.6
\end{tabular}}
\end{table}


In Table~\ref{tab:overall-results}, although both the FOA and MIC datasets are synthesized from the same microphone array, the SELDnet is observed to perform better for FOA than the MIC dataset. This suggests that the spectral and spatial information in the two formats are not identical and methods trained with both the datasets can potentially benefit from mutual information. Finally, we observe that the performance of SELDnet on recordings without polyphony (overlap 1) is significantly better than with polyphony (overlap2). Additionally we can see, at the evaluation set results, that the model does not generalize equally well for different unseen spaces, as it performs better for one of the two rooms (split 7).

\section{Conclusion}
In this paper, we outlined the sound event localization and detection (SELD) task for the DCASE 2020 challenge. In contrast to the DCASE 2019 version of the SELD task which employed static sound scenes, we propose a new dataset with dynamic sound scenes in a reverberant environment. This dataset is synthesized using impulse response trajectories measured at 13-indoor environments. Random segments of these impulse response trajectories are then convolved with isolated sound events from the NIGENs dataset to simulate spatial motion for these sound events. Thereafter, a set of different moving sound events are each placed randomly at various temporal locations of a recording to simulate different polyphony. Lastly, natural ambiance is added to these spatialized and temporalized recordings at different signal-to-noise ratios to make the synthesized sound scene more realistic. To support research on use of various spatial features and recording formats, we provide the proposed dataset in two formats - as first-order Ambisonics, and as a tetrahedral microphone array, both of four-channels. The dataset also provides a fixed evaluation setup for comparing the performances of different participating methods. Finally, we report the baseline scores for the fixed evaluation setup of the dataset with an improved SELDnet method, inspired by the DCASE 2019 challenge submissions.

\bibliographystyle{IEEEtran}
\bibliography{template}
\end{document}